\documentclass[a4paper,11pt]{article}
\usepackage{pos}
\usepackage[T1]{fontenc} 
\usepackage{subcaption}
\usepackage{graphics}

\usepackage{float}
\usepackage{caption}
\usepackage[english]{babel}
\usepackage{adjustbox}

\newcommand{\br}{\begin{eqnarray}}
\newcommand{\er}{\end{eqnarray}}

\def\invfb{\text{fb}^{-1}}

\def\t {\widetilde {t_1}}

\def\C1{\widetilde \chi_1^{\pm}}
\def\mst1 {m_{\t1}}
\def\br {\begin{eqnarray}}
\def\er {\end{eqnarray}}

\def\invfb{\text{fb}^{-1}}

\def \Opq3{\mathcal{O}_{\phi Q}^{(3)}}

\def \Opq1{\mathcal{O}_{\phi Q}^{(1)}}

\def \Ctphi{C_{ t\phi}}
\def \Cphit{C_{\phi t}}
\def \Cpq1{C_{\phi Q}^{(1)}}
\def \Cpq3{C_{\phi Q}^{(3)}}
\def \Cpqm{C_{\phi Q}^{(-)}}
\def \Ctw{C_{tW}}
\mathchardef\mhyphen="2D 

\title{Exploring SMEFT operators through single top-quark production associated with the Higgs boson at the LHC}

\author[a]{Monoranjan Guchait~}
\author*[a]{Arnab Roy}

\affiliation[a]{Department of High Energy Physics,\\
	Tata Institute of Fundamental Research,\\
 Homi Bhabha Road, Mumbai-400005, India}

\emailAdd{guchait@tifr.res.in}
\emailAdd{arnab.roy@tifr.res.in}

\abstract{The Standard Model effective field theory (SMEFT) provides a general framework
to include the dynamics of the beyond standard model physics residing at a certain higher energy scale $\Lambda$. We study the top-quark production along with a Higgs boson and a jet (tHq) at the LHC experiment within the framework of the SMEFT. First, identifying the relevant sensitive dimension-6 SMEFT operators, a strategy is developed to constrain the Wilson Coefficients (WC) corresponding to these associated SMEFT operators using the latest LHC measurements providing a complementary way to the global-fit approach. The best-fit values of these WCs are presented. Finally, we discuss the discovery potential of the signatures of those operators. We find that a discoverable excess due to SMEFT effects can be observed in the tHq process at the LHC with the center of mass energy $\sqrt{s}=13$ TeV and integrated luminosity options ${\cal L}=$300~$\invfb$ and 3000~$\invfb$.}

\FullConference{%
  International Conference on High Energy Physics,\\
  6-13 July, 2022,\\
  Bologna, Italy.
}

\begin{document}
\maketitle

\section{Introduction}
In recent times, the SMEFT framework has received a lot of attention and has become a popular choice to study the effect of UV scale physics
in a more generalized and comprehensive manner~\cite{Buchmuller:1985jz,Grzadkowski:2010es}. 
Plethora of studies are carried out in the literature both phenomenologicallly and experimentally (see~\cite{Guchait:2022ktz}, and references therein). While there can be 59 independent set of SMEFT operators, in this study~\cite{Guchait:2022ktz}, we focus only on a subset of SMEFT operators related to the top quark production in association with a Higgs boson
and a jet,
\br
\rm pp \to tHq.
\label{eq:thq}
\er
This process includes the Yukawa coupling t-t-H and the W-t-b vertex, both having important significance to probe beyond standard model (BSM) physics. Interestingly, some of the relevant operators affecting this process lead to an energy-growing features in scattering amplitudes~\cite{Farina:2012xp}, resulting in excess in the kinematic distributions of some observable. Hence, the prospect of the tHq process in probing this kind of high-scale physics through the SMEFT framework seems to be very promising.
 
 This study is performed in two steps. Firstly, the set of sensitive SMEFT operators affecting the tHq process is identified. It is realized that the same set of SMEFT operators can also affect some other production processes at the LHC involving top quark in the final state. The deviations of various measurements, such as total and differential cross-sections, signal strengths, asymmetries, etc. from the corresponding theoretical predictions of those processes are used to impose constraints on the associated WCs. Secondly, the implications of those constrained operators to various 
kinematic distributions of the tHq process are investigated. Eventually, we demonstrate the discovery potential of the signatures of SMEFT operators presenting the signal significance observed at the tail of the distributions.

\section{SMEFT operators for tHq}
\begin{figure}[htbp]
     \centering
     \begin{subfigure}[b]{0.44\textwidth}
         \centering
         \includegraphics[width=0.45\textwidth]{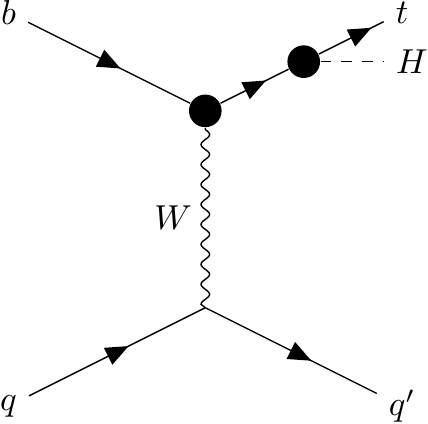}
	\end{subfigure}
     \hfill
     \begin{subfigure}[b]{0.44\textwidth}
         \centering
         \includegraphics[width=0.45\textwidth]{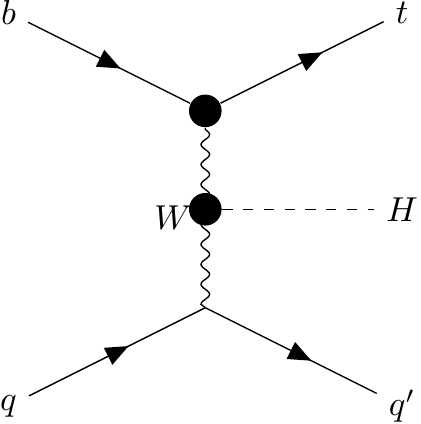}
	\end{subfigure}
	\caption{\small {Feynman diagrams of the tHq process in the t-channel with effective operators affecting $\rm t\mhyphen\bar{t}\mhyphen H$ (left) and H-W-W vertex (right).}}
	\label{fig:feyndiag}
\end{figure}

At first, we identify the set of SMEFT operators associated with the process Eq.~\ref{eq:thq}. The representative Feynman diagrams at the tree level for the above process are shown in Fig.~\ref{fig:feyndiag}, where the blobs represent the vertices expected to be modified by the SMEFT operators.
In this study, we explore the effect of EFT to the t-t-H coupling as well as to the top quark chirality-flipping (W-t-b) interactions through the tHq process. 
On the other hand, following the standard practice, we 
neglect the effects of EFTs to the other vertices not involving top-quark, which can be imposed by invoking  a flavour symmetry requirement~\cite{Guchait:2022ktz}.
Under this flavor assumption, a larger set of SMEFT operators reduces to seven most relevant operators for the tHq process as shown in Table~\ref{tab:op_prod}, where symbols have their usual meaning.

\begin{table}[htbp]
	\caption{\small {SMEFT operators contributing to the $\rm tHq$ production (Fig.\ref{fig:feyndiag}) in the Warsaw basis, with a symmetry assumption~\cite{Guchait:2022ktz}.}}
	\centering 
	\resizebox{11cm}{!}{
	\begin{tabular}{c c c c c}	
		\hline 
		Operator & Coefficient   & Definition & Coupling & Sensitive process \\
		\hline
		
		$\mathcal{O}_{ t\phi}$ & $C_{t\phi}$ & $(\phi^{\dagger}\phi-v^2/2)\bar{Q}t\tilde{\phi}$& $\rm t\mhyphen\bar{t}\mhyphen H$ & $\rm t\bar{t}H$, tHq \\
		
		$\mathcal{O}_{\phi t}$ & $C_{\phi t}$& $i(\phi^{\dagger}\overleftrightarrow{D_{\mu}}\phi)\bar{t}\gamma^{\mu}t$ &  $\rm t\mhyphen\bar{t}\mhyphen H$, $\rm t\mhyphen \bar{t}\mhyphen V$ & $\rm t\bar{t}H$, tHq, tj, tV, $\rm t\bar{t}Z$ \\
		$\mathcal{O}_{\phi Q}^{(1)}$ & $C_{\phi Q}^{(1)}$ & $i(\phi^{\dagger}\overleftrightarrow{D_{\mu}}\phi)\bar{Q}\gamma^{\mu}Q$& $\rm t\mhyphen\bar{t}\mhyphen H$,W-t-b & $\rm t\bar{t}H$, tHq, tj, tV, $\rm t\bar{t}V$\\
		
		$\mathcal{O}_{\phi Q}^{(3)}$ &$C_{\phi Q}^{(3)}$ &$i(\phi^{\dagger}\overleftrightarrow{D_{\mu}}\tau_{I}\phi)\bar{Q}\gamma^{\mu}\tau^IQ$ & $\rm t\mhyphen\bar{t}\mhyphen H$,W-t-b & $\rm t\bar{t}H$, tHq, tj, tV, $\rm t\bar{t}V$\\
		
		$\mathcal{O}_{tW}$ & $C_{tW}$ & $i(\bar{Q}\sigma^{\mu\nu}\tau_It)\tilde{\phi}W^I_{\mu\nu}$& W-t-b & $\rm t\bar{t}H$, tHq, tj, tV, $\rm t\bar{t}V$\\
		$\mathcal{O}_{\phi W}$ & $C_{\phi W}$ & $(\phi^{\dagger}\phi-v^2/2)W_{\mu\nu}^{I} W^I_{\mu\nu}$& H-W-W &   tHq\\
		$\mathcal{O}_{\phi D}$ & $C_{\phi D}$ & $(\phi^{\dagger}{D_{\mu}}\phi)(\phi^{\dagger}{D^{\mu}}\phi)$& H-W-W &  tHq\\
		\hline
	\end{tabular}
	}
	\label{tab:op_prod}
\end{table} 

Notice that, the 
$\mathcal{O}_{\phi W}$ and $\mathcal{O}_{\phi D}$ are tightly constrained by the measurements of the electroweak precision observables (EWPO), and found insensitive to the production cross-section nor or the kinematic distributions with the allowed ranges. So $\mathcal{O}_{\phi W}$ and $\mathcal{O}_{\phi D}$ are not taken into account in our analysis~\cite{Guchait:2022ktz}. For similar reasons, the effects of SMEFT operators on the decay vertex of Higgs and vector bosons are not counted in our analysis. On the contrary, EFT effects on the top quark decay vertex (W-t-b) are taken into consideration as it is expected to provide a similar effect as in the production vertex. Notably, instead of using $\mathcal{O}_{\phi Q}^{(1)}$ directly, a combination of $\mathcal{O}_{\phi Q}^{(1)}$ and 
$\mathcal{O}_{\phi Q}^{(3)}$, as defined by,
\br
\mathcal{O}_{\phi Q}^{(-)}=\mathcal{O}_{\phi Q}^{(1)}-\mathcal{O}_{\phi Q}^{(3)},
\er
is used in this study. Finally, having all these considerations, we end up with the following five relevant set of operators for our present study,  
\br
\mathcal{O}_{ t\phi},\mathcal{O}_{\phi t},\mathcal{O}_{\phi Q}^{(-)},\mathcal{O}_{\phi Q}^{(3)},\mathcal{O}_{tW}.
\er
\section{Constraining operators $\mathcal{O}_{ t\phi},\mathcal{O}_{\phi t},\mathcal{O}_{\phi Q}^{(-)},\mathcal{O}_{\phi Q}^{(3)},\mathcal{O}_{tW}$}

Note that, the cross-section (or any related observable) of any process can be expressed in a more compact form as,
\br
\sigma^{EFT}({C})=\sigma^{SM}+\sum_{i=1}^{n} C_i\beta_i+\sum_{j\leq k}^{n}C_jC_k\gamma_{jk}.
\label{eq:sigma}
\er
The numerical values of these coefficients are obtained by fitting the variation of cross-sections, evaluated using \nolinkurl{MG5aMC_atNLO}~\cite{Alwall:2014hca} interfacing with \nolinkurl{SMEFTatNLO}~\cite{Degrande:2020evl} feynrules UFO,  with an appropriate polynomial considering WCs as variables while $\beta_i, \gamma_{ij}$ as fitting parameters.

In order to have a quantitative assessment of the level sensitivity
of these SMEFT operators to various processes,
we resort to the formalism of the Fisher information matrix (FIM)~\cite{Brehmer:2016nyr}.
The FIM ($I_{ij}$) for all our considered processes and operators are derived and presented in Table~\ref{tab:fishinfo}, where each column of this table presents the relative sensitivities of the set of WCs corresponding to respective physics processes, normalised as $\mathrm{\sum_{i=1}^{N_{WC}} |diag}(I_{ii})|=100$~\cite{Guchait:2022ktz}.

\begin{table}[htbp]
	\caption{\small {The relative sensitivity of WCs corresponding to various processes normalized to the scale of 100.}}
	\centering
	\resizebox{6.5cm}{!}{
	\begin{tabular}{c c c| c| c| c| c| c}
		\hline
		& WC &tH&$\rm t\bar{t}H$ &tj &$\rm t\bar{t}V$ &tZ &  tW\\
		\hline
		&$\Cpq3$&15.7 & 2.4 & 79.6 & 2.3 & 50.1& 64.0\\
		&$\Cpqm$&15.6 &0.4 &0.4 &55.3 &37.2& 0.7\\
		&$\Cphit$& 16.4 &3.7 &0.4 &41.6 &0.2& 0.1\\
		&$\Ctphi$ &33.4 &92.1 &0.8 &0.6 &0.003& 0.002\\
		&$\Ctw$& 18.5 &1.2 &18.4 &0.4 &12.4& 35.1\\
		\hline
	\end{tabular}
	}
	\label{tab:fishinfo}
\end{table}

\begin{table}[htbp]
	\caption{\small {Summary of number of measurements used from various relevant processes~\cite{Guchait:2022ktz}.}}
	\centering
	\resizebox{7.5cm}{!}{
	\begin{tabular}{c c c c c |c}
		\hline
		Process & $t\bar{t} H,tH$ & $tj$ & $t\bar{t}Z,t\bar{t}W$ & $tZ,tW$ & Total\\
		\hline
		No. of data & 8 & 13 & 17 & 10 & 48\\
		\hline
	\end{tabular}
	}
	\label{tab:data}
\end{table}

In Table~\ref{tab:data}, a summary of measurements used to constrain the WCs are presented~\cite{Guchait:2022ktz}. We find the constrained values of these five WCs by the method of $\chi^2$-minimization, performed  
using the framework of TMinuit with the option MIGRAD~\cite{James:1994vla}. Depending on the order of the used theoretical values, the fitting is performed both at linear and quadratic level (as shown in Eq~\ref{eq:sigma}). The best-fit values for the combined quadratic fit for the respective WCs are presented in Table \ref{tab:bestfit}.
\begin{table}[htbp]
	\caption{\small {Best-fit values of the WCs, referred together as `SM+EFT'.}}
	\centering
		\resizebox{9.5cm}{!}{
	\begin{tabular}{cccccc}
		\hline  
		&  $\mathcal{O}_{t\phi}$ & $\mathcal{O}_{\phi Q}^{(3)}$ & $\mathcal{O}_{\phi t}$ & $\mathcal{O}_{\phi Q}^{(-)}$ & $\mathcal{O}_{tW}$ \\
		\hline
		Best-fit values& $-0.2^{+1.8}_{-1.6}$ & $1.5^{+0.5}_{-0.8}$  &$1.9 ^{+1.9}_{-10.0}$  & $-3.7^{+4.1}_{-1.3}$& $-1.2^{+1.0}_{-0.4}$  \\
		\hline
	\end{tabular}
	}
	\label{tab:bestfit}
\end{table} 

\section{Implications at the LHC}
The impact of the SMEFT operators on any process is expected to be observed at the tail of the kinematic distribution of certain chosen observables constructed out of the momenta of final state particles of the tHq process. In this case, the primary reason for such possible deviations can be attributed to the energy growth in the scattering sub-amplitude $\rm bW\to tH$~\cite{Farina:2012xp}.

We perform a detailed analysis by splitting the phase space into two regions, 
(a) boosted region, with  $\rm p_T (H)>300~GeV$, and (b) 
non-boosted region, with $\rm p_T (H)<300~GeV$. 
Additionally, both the hadronic and the leptonic decay channels of top-quark are 
considered separately, requiring the following final state,
\br
\rm H_{ reco}+ t_{ reco}+ n\mhyphen jets~~~~(hadronic~ final~ state)
\label{eq:signal1}\\
\rm ~~~~~~H_{ reco}+\ell+ n_b\mhyphen jets + n\mhyphen jets ~~~~(semi-leptonic~ final~ state),
\label{eq:signal2}
\er
with $n_b,n \geq 1$. 	
The dominant SM background contributions are expected to be due to the following processes,
\br
\rm p~p \rightarrow t\bar{t},~ t\bar{t}H,~ t\bar{t}Z,~ t\bar{t}b\bar{b},~ t\bar{t}W, WH.
\label{eq:backgrounds}
\er
We first estimate the level of contamination due to the SM backgrounds (Eq.~\ref{eq:backgrounds}) at the signal region, and then also checked the same including the EFT effects~\cite{Guchait:2022ktz}.

\begin{figure}[htbp]
	\begin{subfigure}[b]{0.5\textwidth}
		\centering
		\includegraphics[width=6.5 cm]{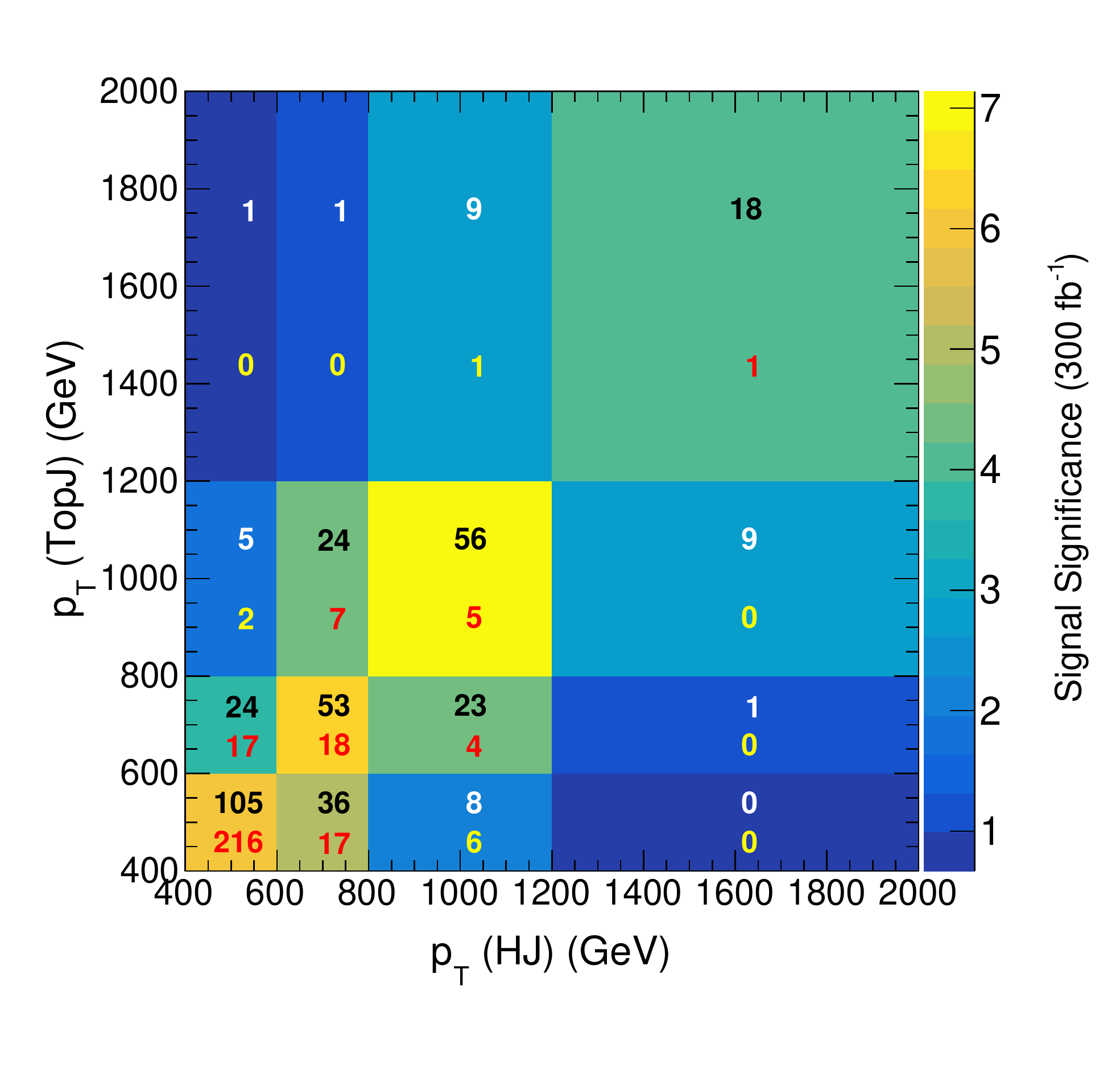}
	\end{subfigure}
	\begin{subfigure}[b]{0.5\textwidth}
		\centering
		\includegraphics[width=6.5 cm]{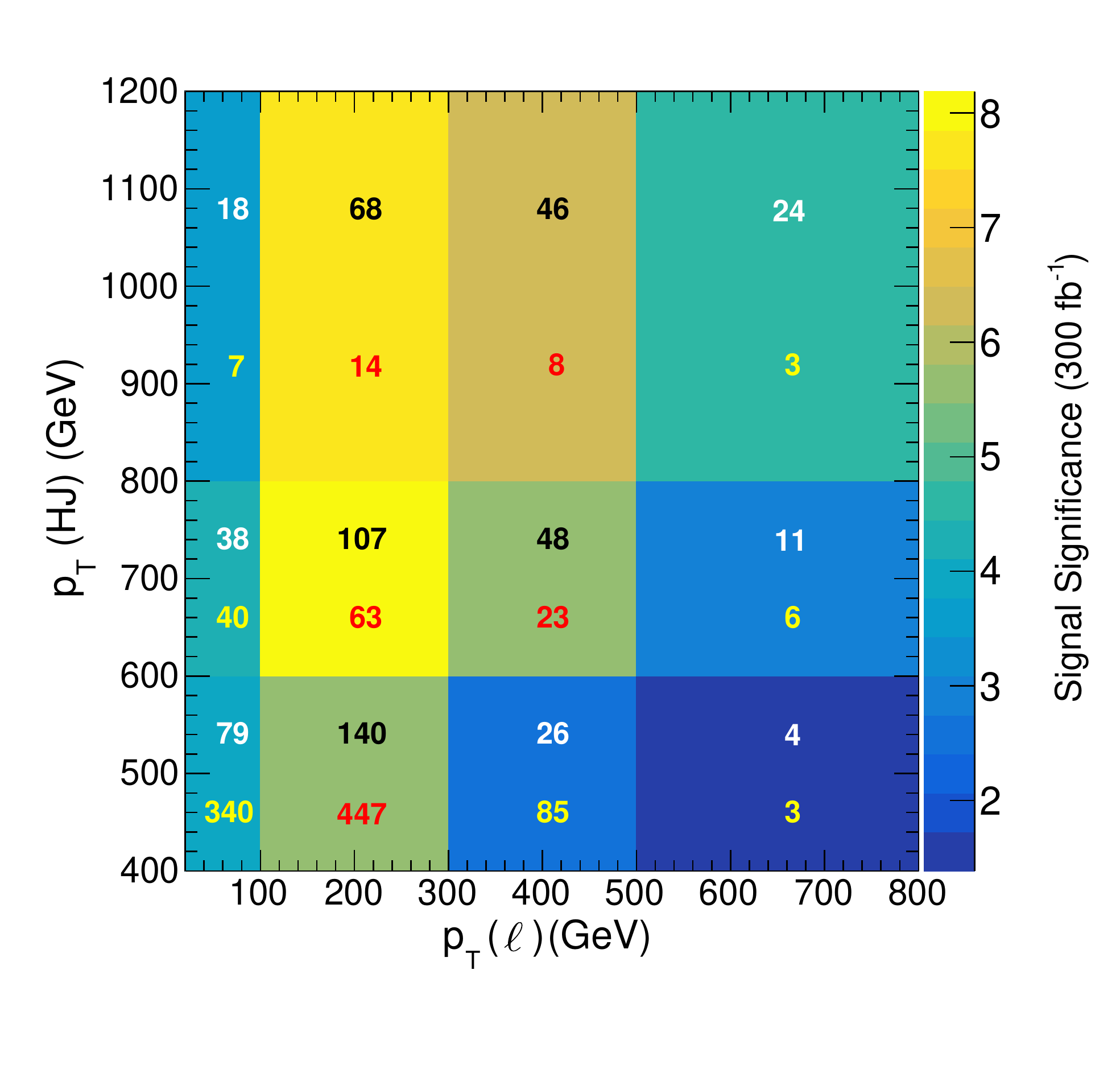}
	\end{subfigure}
	\caption{\small {2D distribution of $\rm p_T(HJ)$ and $\rm p_T(TopJ)$ in terms of signal significance ($\rm S/\sqrt{(S+B)}$) at $\rm \mathcal{L}=300~fb^{-1}$ for hadronic (left) and leptonic (right) final state. The upper (lower) entries present the number of event yields for signal (total background).}}
	\label{fig:sens_2d}
\end{figure}

Initially, the effects of the SMEFT operators in various kinematic distributions, such as the transverse momentum of reconstructed Higgs boson, top etc. and the invariant mass of Higgs-top and lepton-Higgs, is studied. It is observed that, for the best-fit values of the WCs (Table~\ref{tab:bestfit}), a countable excess of events can be observed for a luminosity $\rm \mathcal{L}=300~fb^{-1}$ in the boosted region of the tHq production process. The feature of the `high scale sensitivity' reflected in the excess can be 
captured in the `bin by bin' estimation of the respective kinematic variable, rather than
counting a total number of events. Basically, the region where excess is observed is divided into several bins, and the number of events corresponding to those for both the signal and total background is counted. In Fig.~\ref{fig:sens_2d} we present the signal region in 2D planes dividing 
into the $\rm p_T$ bins of the HJ and TopJ for hadronic(left) and leptonic(right)
case. In each bin, the upper (lower) entries present the signal (total background)
yields. The respective colors of each bin show the level of significances,
$\rm S/\sqrt{(S+B)}$, for a luminosity option $\rm \mathcal{L}=300~fb^{-1}$.

\section{Summary}
We present a detailed study of 
new physics effects in the single top quark production associated with the SM Higgs
boson (tHq) within the SMEFT framework. The constrained ranges of chosen SMEFT operators 
are obtained following the technique of $\chi^2$-minimization using most sensitive experimental measurements. In order to study the effects of the SMEFT operators on the kinematic distributions, we perform a dedicated simulation of tHq process at the LHC energy, $\sqrt{s}=13$ TeV, for both the hadronic and leptonic channels according to the decay of top quark. A $\rm p_T$-binned analysis turns out to be very effective to single out excesses at the tails of the distributions. It is seen that, a reasonable signal significance $\sim 5 \sigma$ can be achieved in most of the higher $\rm p_T$-bins at the high luminosity option $\rm \mathcal{L}=300~fb^{-1}$, presumably which will further increase at HL-LHC option $\rm \mathcal{L}=3000~fb^{-1}$. In conclusion, our analysis shows that, the scope of the tHq process in boosted scenario is very promising in order to observe discoverable excess due to EFT effects.

{\bf Acknowledgements:} AR is thankful to `Infosys-TIFR Leading Edge Travel Grant’ for providing support to present this study at the ICHEP, 2022, Bologna, Italy.
\bibliographystyle{JHEP}
\bibliography{eft}

\providecommand{\href}[2]{#2}\begingroup\raggedright\begin{thebibliography}{1}

\bibitem{Buchmuller:1985jz}
W.~Buchmuller and D.~Wyler, \emph{{Effective Lagrangian Analysis of New
  Interactions and Flavor Conservation}},
  \href{https://doi.org/10.1016/0550-3213(86)90262-2}{\emph{Nucl. Phys. B}
  {\bfseries 268} (1986) 621}.

\bibitem{Grzadkowski:2010es}
B.~Grzadkowski, M.~Iskrzynski, M.~Misiak and J.~Rosiek, \emph{{Dimension-Six
  Terms in the Standard Model Lagrangian}},
  \href{https://doi.org/10.1007/JHEP10(2010)085}{\emph{JHEP} {\bfseries 10}
  (2010) 085} [\href{https://arxiv.org/abs/1008.4884}{{\ttfamily 1008.4884}}].

\bibitem{Guchait:2022ktz}
M.~Guchait and A.~Roy, \emph{{Exploring SMEFT operators in the tHq production
  at the LHC}},  \href{https://arxiv.org/abs/2210.05503}{{\ttfamily
  2210.05503}}.

\bibitem{Farina:2012xp}
M.~Farina, C.~Grojean, F.~Maltoni, E.~Salvioni and A.~Thamm, \emph{{Lifting
  degeneracies in Higgs couplings using single top production in association
  with a Higgs boson}},
  \href{https://doi.org/10.1007/JHEP05(2013)022}{\emph{JHEP} {\bfseries 05}
  (2013) 022} [\href{https://arxiv.org/abs/1211.3736}{{\ttfamily 1211.3736}}].

\bibitem{Alwall:2014hca}
J.~Alwall, R.~Frederix, S.~Frixione, V.~Hirschi, F.~Maltoni, O.~Mattelaer
  et~al., \emph{{The automated computation of tree-level and next-to-leading
  order differential cross sections, and their matching to parton shower
  simulations}}, \href{https://doi.org/10.1007/JHEP07(2014)079}{\emph{JHEP}
  {\bfseries 07} (2014) 079} [\href{https://arxiv.org/abs/1405.0301}{{\ttfamily
  1405.0301}}].

\bibitem{Degrande:2020evl}
C.~Degrande, G.~Durieux, F.~Maltoni, K.~Mimasu, E.~Vryonidou and C.~Zhang,
  \emph{{Automated one-loop computations in the standard model effective field
  theory}}, \href{https://doi.org/10.1103/PhysRevD.103.096024}{\emph{Phys. Rev.
  D} {\bfseries 103} (2021) 096024}
  [\href{https://arxiv.org/abs/2008.11743}{{\ttfamily 2008.11743}}].

\bibitem{Brehmer:2016nyr}
J.~Brehmer, K.~Cranmer, F.~Kling and T.~Plehn, \emph{{Better Higgs boson
  measurements through information geometry}},
  \href{https://doi.org/10.1103/PhysRevD.95.073002}{\emph{Phys. Rev. D}
  {\bfseries 95} (2017) 073002}
  [\href{https://arxiv.org/abs/1612.05261}{{\ttfamily 1612.05261}}].

\bibitem{James:1994vla}
F.~James, ``{MINUIT Function Minimization and Error Analysis: Reference Manual
  Version 94.1}.'' 1994.

\end{thebibliography}\endgroup

\end{document}